\documentclass{nature}
\usepackage{epsfig}
\usepackage{amsmath}

\begin{document}

\title{Maximal Holevo quantity based on weak measurements}
\author{Yao-Kun Wang$^{1,2}$, Shao-Ming Fei$^{3,4}$, Zhi-Xi Wang$^{3}$, Jun-Peng Cao$^{1,5}$ and Heng
Fan$^{1,5,\star}$}
\maketitle

\begin{affiliations}
\item
Beijing National Laboratory for Condensed Matter Physics, Institute
of Physics, Chinese Academy of Sciences, Beijing 100190, China
\item
College of Mathematics,  Tonghua Normal University,
 Tonghua, Jilin 134001, China
 \item
 School of Mathematical Sciences,  Capital Normal
University,  Beijing 100048,  China
\item
Max-Planck-Institute for Mathematics in the Sciences, 04103 Leipzig, Germany
\item
Collaborative Innovative Center of Quantum Matter, Beijing 100190, China

$^\star$e-mail: hfan@iphy.ac.cn.

\end{affiliations}

\begin{abstract}
The Holevo bound is a keystone in many applications of quantum information theory.
We propose ``weak maximal Holevo quantity'' with weak measurements as the generalization of the standard Holevo quantity
which is defined as the optimal projective measurements. The scenarios that weak measurements is necessary are
that only the weak measurements can be performed because for example the system is macroscopic or
that one intentionally tries to do so such that the disturbance on the measured system
can be controlled for example in quantum key distribution protocols.
We evaluate systematically the weak maximal Holevo quantity for Bell-diagonal states and
find a series of results. Furthermore, we find that weak measurements can be realized by noise and project measurements.
\end{abstract}

Weak measurements was introduced by Aharonov, Albert, and Vaidman (AAV) \cite{Aharonov} in 1988.
The standard measurements can be realized as a sequence of weak measurements which result in small changes to the quantum state for all outcomes \cite{Oreshkov}.
Weak measurements realized by some experiments is also very useful for high-precision measurements\cite{Hosten,resch,Dixon,howell,Gillett}.

The quantum correlations of quantum states include entanglement and other kinds of nonclassical correlations.
It is well known that the quantum correlations are more general than the well-studied entanglement \cite{bennett, zurek1}.
Quantum discord, a quantum correlation measure differing from entanglement,
is introduced by Oliver and Zurek\cite{Ollivier} and independently
by Henderson and Vedral\cite{henderson}.
It quantifies the difference between the mutual information and maximum classical mutual information,
i.e., it is a measure of the difference between total correlation and the classical correlation.
Significant developments have been achieved in studying properties and applications of quantum discord.
In particular, there are some analytical expressions for quantum discord for two-qubit states, such as
 for the $X$ states \cite{luo,Ali,Li,chen,shi,Vinjanampathy}.
Besides, researches on the dynamics of quantum discord in various noisy environments have revealed many attractive features \cite{werlang,bwang,auccaise}.
 It is demonstrated that discord is more robust than entanglement for both Markovian and non-Markovian dissipative processes.
 As with projection measurements, weak measurements are also applied to study the quantification of quantum correlation.
For example, the super quantum correlation based on weak measurements has attracted much attention \cite{singh,wang,li,singh2,hu}.

 In general, maximum classical mutual information is  called classical correlation which represents the difference in Von Neumann entropy before and after the measurements\cite{henderson}.
 A similarly defined quantity is the Holevo bound which measures the capacity of quantum states for classical communication\cite{holevo,bena}.
 The Holevo bound is an exceedingly useful upper bound on the accessible information that plays an important role in many applications of quantum information theory\cite{nielsen}.
 It is a keystone in the proof of many results in quantum information theory\cite{lupo,zhang,lloyd,roga,wu}.

The maximal Holevo quantity and classical correlation are both classical and based on von Neumann measurements.
Due to the fundamental role of weak measurements, it is interesting to know how those classical correlations will be if weak measurements are taken into account.
Recently, it is shown that weak measurements performed on one of the subsystems can lead to ``super quantum discord'' which is always larger than the normal quantum discord captured by projective measurements \cite{singh}.
It is natural to ask whether weak measurements can also capture more classical correlations.
In this article, we shall give the definition of ``super classical correlation'' by weak measurements as the generalization of classical correlation defined for standard projective measurements.
As the generalization of the maximal Holevo quantity defined for projective measurements, we propose ``weak maximal Holevo quantity'' according weak measurements.
Interestingly, by tuning continuously from strong measurements to weak measurements, the discrepancy between the weak maximal Holevo quantity and the maximal Holevo quantity becomes larger.
Such phenomenon also exits between super classical correlation and classical correlation.
In comparison with super quantum discord which is larger than the standard discord,
the weak maximal Holevo quantity and super classical correlation becomes less when weak measurements are applied, while they are completely the same for projective measurements.
In this sense, weak measurements do not capture more classical correlations.
It depends on the specified measure of correlations.
We calculate the maximal Holevo quantity for Bell-diagonal states, and compare the results with classical correlation.
We give super classical correlation and  weak maximal Holevo quantity for Bell-diagonal states and compare the relations among super quantum correlations, quantum correlations, classical correlation and super classical correlation and entanglement.
The dynamic behavior of  weak maximal Holevo quantity under decoherence is also investigated.

\section*{Results}

\subsection{ Maximal Holevo quantity and weak maximal Holevo quantity.}

The quantum discord for a bipartite quantum state $\rho_{AB}$ with the projection measurements $\{\Pi^B_i\}$ performed on the subsystem $B$ is the difference between the mutual information $I(\rho_{AB})$ \cite{partovi} and classical correlation $J_B(\rho_{AB})$\cite{henderson}:
\begin{equation}
D(\rho_{AB})=I(\rho_{AB})-J_B(\rho_{AB}),\nonumber
\end{equation}
where
\begin{equation}
I(\rho_{AB})=S(\rho_{A})+S(\rho_{B})-S(\rho_{AB}),\nonumber
\end{equation}
\begin{eqnarray}
J_B(\rho_{AB})&=&\sup_{\{B_k\}}\{S(\rho_{A})-\sum_i p_i S(\rho_{A|i})\}\nonumber \\
&=&S(\rho_{A})-\min_{\{\Pi_i^B\}} \sum_i p_i S(\rho_{A|i})\nonumber
\end{eqnarray}
with the minimization going over all projection measurements $\{\Pi^B_i\}$,
where $S(\rho) = - \mbox{tr} \left(\rho \log_2 \rho\right)$ is the von Neumann entropy of a quantum state $\rho$, $\rho_A$, $\rho_B$ are the reduced density matrices of $\rho_{AB}$ and
\begin{equation}
p_i = \mbox{tr}_{AB}[(I_A \otimes \Pi^B_i ) \rho_{AB} ( {I}_A \otimes \Pi^B_i) ],\
\rho_{A|i} = \frac{1}{p_i} \mbox{tr}_B[({I}_A \otimes \Pi^B_i) \rho_{AB} ({I}_A \otimes \Pi^B_i)].\nonumber
\end{equation}

The Holevo quantity of the ensemble $\{p_{i};\rho_{A|i}\}$ \cite{wu} that is
prepared for A by B via its local measurements is given by
\begin{eqnarray}
\chi\{\rho_{AB}|\{\Pi^B_i\}\}=\chi\{p_{i};\rho_{A|i}\}\equiv S(\sum_{i}p_{i}\rho_{A|i})-\sum_{i}p_{i}S(\rho_{A|i}).
\end{eqnarray}
It denotes the upper bound of A's accessible information about B's measurement result when B projects its system by the projection operaters $\{\Pi^B_i\}$.
The classical correlation in the state $\rho_{AB}$ is defined as the maximal Holevo quantity\cite{wu} over all local projective measurements on B's system:
\begin{equation}
C_{1}(\rho_{AB}) \equiv \max_{\{\Pi^B_i\}}\chi\{\rho_{AB}|\Pi^B_i\}\}.
\end{equation}


The weak measurement operators are given by \cite{Oreshkov}
\begin{eqnarray}
 P(x) &=& \sqrt{\frac{(1-\tanh x)}{2}} \Pi_0 + \sqrt{ \frac{(1+\tanh x)}{2}} \Pi_1,  \nonumber\\
 P(-x) & = & \sqrt{\frac{(1+\tanh x)}{2}}\Pi_0 + \sqrt{\frac{(1-\tanh x)}{2}}\Pi_1, \nonumber\\
\end{eqnarray}
where $x$ is the measurement strength parameter, $\Pi_0$ and $\Pi_1$ are two orthogonal projectors with $\Pi_0 + \Pi_1 =I$. The weak measurements operators satisfy: (i) $P^{\dagger}(x)P(x) + P^{\dagger}(-x)P(-x) = I$, (ii) $\lim_{x \rightarrow \infty} P(x) = \Pi_0$ and  $\lim_{x \rightarrow \infty} P(-x) = \Pi_1$.

Recently, super quantum discord for bipartite quantum state \(\rho_{AB}\) with weak measurements on the subsystem $B$ has been proposed \cite{singh}.
Similarly to the definition of quantum discord, we give the another form of definition of super quantum discord.
We define super classical correlation $J_B^{w}(\rho_{AB})$ for bipartite quantum state $\rho_{AB}$ with the weak measurements $\{P^{B}(\pm x)\}$ performed on the subsystem $B$ as follow.
The super quantum discord denoted by $D_w(\rho_{AB})$ is the difference between the mutual information $I(\rho_{AB})$ and super classical correlation $J_B^{w}(\rho_{AB})$, i.e.,
\begin{equation}
D_w(\rho_{AB})=I(\rho_{AB})-J_B^{w}(\rho_{AB})\nonumber,
\end{equation}
where
\begin{equation}
I(\rho_{AB})=S(\rho_{A})+S(\rho_{B})-S(\rho_{AB}),\nonumber
\end{equation}
\begin{eqnarray}
J_B^{w}(\rho_{AB})&=&\sup_{\{B_k\}}\{S(\rho_{A})-S_w(A|\{P^{B}(x)\})\}\nonumber \\
&=&S(\rho_{A})-\min_{\{P(\pm x)\}} \{p(x) S(\rho_{A|P^{B}(x)}) + p(-x) S(\rho_{A|P^{B}(-x)})\},
\end{eqnarray}
with the minimization going over all projection measurements \(\{\Pi^B_i\}\),
\begin{equation}
 S_w(A|\{P^{B}(x)\})= p(x) S(\rho_{A|P^{B}(x)}) + p(-x) S(\rho_{A|P^{B}(-x)}),\nonumber
\end{equation}
\begin{equation}
 p(\pm x) =\mbox{tr}_{AB}[(I \otimes P^{B}(\pm x)) \rho_{AB} (I \otimes P^{B}(\pm x))],\label{probab}
\end{equation}
\begin{equation}
 \rho_{A|P^{B}(\pm x)}=\frac{\mbox{tr}_{B}[(I \otimes P^{B}(\pm x)) \rho_{AB} (I \otimes  P^{B}(\pm x))]}
{\mbox{tr}_{AB}[(I \otimes P^{B}(\pm x)) \rho_{AB} (I \otimes P^{B}(\pm x))]},\label{state1}
\end{equation}
$\{P^{B}(x)\}$ is weak measurement operators performed on the subsystem $B$.

Now, let us define the weak Holevo quantity of the ensemble $\{p(\pm x); \rho_{A|P^{B}(\pm x)}\}$ based on weak measurements on the subsystem $B$,
\begin{eqnarray}
\chi^{w}\{\rho_{AB}|\{P(\pm x)\}\}&=&\chi\{p(\pm x);\rho_{A|P^{B}(\pm x)}\}\nonumber\\
&=& S\left(\sum_{\pm x}p(\pm x)\rho_{A|P^{B}(\pm x)}\right)-\sum_{\pm x}p(\pm x)S\left(\rho_{A|P^{B}(\pm x)}\right).
\end{eqnarray}
It denotes the upper bound of A's accessible information about B's measurements results when B projects the system with the weak measurements operaters $\{P(\pm x)\}$.
The weak maximal Holevo quantity over all local weak measurements on B's system is given by:
\begin{equation}
C_{1}^{w}(\rho_{AB})= \max_{\{P(\pm x)\}}\chi^{w}\{\rho_{AB}|\{P(\pm x)\}\}.
\end{equation}

Next, we consider the maximal Holevo quantity and weak maximal Holevo quantity for two-qubit Bell-diagonal states,
\begin{eqnarray}
\rho_{AB}=\frac{1}{4}(I\otimes I+\sum_{i=1}^3c_i\sigma_i\otimes\sigma_i),
\end{eqnarray}
where  $I$ is the identity matrix, $-1\le c_i \le 1$. The marginal states of $\rho_{AB}$ are $\rho_{A}=\rho_{B}=\frac{I}{2}$.

The maximal Holevo quantity for Bell-diagonal states is given as
\begin{eqnarray}\label{cc}
C_{1}(\rho_{AB})&=&\max_{\{\Pi^B_i\}}\chi\{\rho_{AB}|\Pi^B_i\}\}.\nonumber\\
&=&\frac{1-C}{2}\log(1-C)+\frac{1+C}{2}\log(1+C),
\end{eqnarray}
where $C=\max\{|c_{1}|, |c_{2}|, |c_{3}|\}$.
We find that the maximal Holevo quantity $C_{1}(\rho_{AB})$ equals to the classical correlation $J_B(\rho_{AB})$,
\begin{eqnarray}
C_{1}(\rho_{AB})=J_B(\rho_{AB}).
\end{eqnarray}

The weak maximal Holevo quantity of two-qubit Bell-diagonal states is given by
\begin{eqnarray}\label{sc}
C_{1}^{w}(\rho_{AB})&=& \max_{\{P(\pm x)\}}\chi^{w}\{\rho_{AB}|\{P(\pm x)\}\}\nonumber\\
&=&\frac{1-C\tanh x}{2}\log(1-C\tanh x)+\frac{1+C\tanh x}{2}\log(1+C\tanh x).
\end{eqnarray}

The super classical correlation of two-qubit Bell-diagonal states is given by
\begin{eqnarray}\label{sjc}
J_B^{w}(\rho_{AB})&=&\sup_{\{B_k\}}\{S(\rho_{A})-S_w(A|\{P^{B}(x)\})\} \nonumber\\
&=&\frac{1-C\tanh x}{2}\log(1-C\tanh x)+\frac{1+C\tanh x}{2}\log(1+C\tanh x).
\end{eqnarray}

The weak maximal Holevo quantity $C_{1}^{w}(\rho_{AB})$ equals to  the super classical correlation $J_B^{w}(\rho_{AB})$, i.e.,
\begin{eqnarray}
C_{1}^{w}(\rho_{AB})=J_B^{w}(\rho_{AB}).
\end{eqnarray}

Next, we compare the weak maximal Holevo quantity(super classical correlation), the maximal Holevo quantity(classical correlation), super quantum discord, quantum discord, and entanglement of formation. For simplicity, we choose Werner states, $c_{1}=c_{2}=c_{3}=-z$,
\begin{equation}
\rho_{AB} = z |\Psi^-\rangle\langle\Psi^-|+\frac{(1-z)}{4}I, z\in[0,1],
\end{equation}
where $|\Psi^-\rangle=(|01\rangle - |10\rangle)/\sqrt{2}$. Set $z=\frac{\alpha}{2-\alpha}$. The Werner states have the form
\begin{equation}
\rho_{w}=\frac{1}{2(2-\alpha)}\left(I-\alpha P\right) ,
\label{Wernerstates}
\end{equation}
where $-1\leq\alpha\leq1$, $I$ is the identity operator in the $4$-dimensional Hilbert space, and $P=\sum_{i,j=1}^{2}\left|i\right\rangle \left\langle j\right|\otimes\left|j\right\rangle \left\langle i\right|$ is the operator that exchanges A and B.
The entanglement of formation $E_{f}$ for the Werner states is given as $E_{f}(\rho_{w})=h\left(\frac{1}{2}(1+\sqrt{1-[\max(0,\frac{2\alpha-1}{2-\alpha})]^{2}})\right)$,
by $h(x)\equiv - x\log_{2} x- (1-x) \log_{2}(1-x)$.

The maximal Holevo quantity for werner states is given by, see Eq. (\ref{cc}) in section Method,
\begin{eqnarray}
C_{1}(\rho_{AB})=\frac{1-z}{2}\log(1-z)+\frac{1+z}{2}\log(1+z).
\end{eqnarray}

The weak maximal Holevo quantity for werner states is given by, see Eq. (\ref{sc}) in section Method,
\begin{eqnarray}
C_{1}^{w}(\rho_{AB})=\frac{1-z\tanh x}{2}\log(1-z\tanh x)+\frac{1+z\tanh x}{2}\log(1+z\tanh x).
\end{eqnarray}

The quantum discord for Werner states is given by \cite{luo}
\begin{eqnarray}
D(\rho_{AB})=\frac{1-z}{4}\log(1-z)-\frac{1+z}{2}\log(1+z)+\frac{1+3z}{4}\log(1+3z).
\end{eqnarray}

And the super quantum discord for Werner states is given by \cite{singh}
\begin{eqnarray}
D_w(\rho_{AB}) &=&\frac{3(1-z)}{4}\log\left(\frac{1-z}{4}\right)+\frac{(1+3z)}{4}\log\left(\frac{1+3z}{4}\right)\nonumber\\
& &+1-[\frac{(1-z\tanh x)}{2}\log\left(\frac{1-z\tanh x}{2}\right)\nonumber\\
& &+\frac{(1+z\tanh x)}{2}\log\left(\frac{1+z\tanh x}{2}\right)].
\end{eqnarray}

In Fig.1 we plot the weak maximal Holevo quantity, the maximal Holevo quantity, super quantum discord, quantum discord, and entanglement of formation for the Werner state.
We find that super quantum discord , quantum discord, the maximal Holevo quantity and the weak maximal Holevo quantity have the relations, $D_w\geq D>J_{B}(C_{1})\geq J_{B}^{w}(C_{1}^{w})$.
For the case of projection measurements, $\lim x \rightarrow \infty $, we have $D_{w}=D, J_{B}(C_{1})= J_{B}^{w}(C_{1}^{w})$. The weak maximal Holevo quantity approaches to zero for smaller values of $x$.
The weak maximal Holevo quantity approaches to the maximal Holevo quantity and super quantum discord approaches to quantum discord for larger values of $x$.
The weak maximal Holevo quantity and the maximal Holevo quantity are larger than the entanglement of formation for small $z$ and smaller than the entanglement of formation for big $z$.
It shows that the weak maximal Holevo quantity and the maximal Holevo quantity can not always capture more correlation than the entanglement as super quantum discord and quantum discord do.

As a natural generalization of the classical mutual information, the classical correlation represents the difference in Von Neumann entropy before and after projection measurements, i.e.,
\begin{eqnarray}
J_B(\rho_{AB})=S(\rho_{A})-\min_{\{\Pi_i^B\}} \sum_i p_i S(\rho_{A|i}).\nonumber
\end{eqnarray}

Similarly, the super classical correlation represents the difference in Von Neumann entropy before and after weak measurements, i.e.,
\begin{eqnarray}
J_B^{w}(\rho_{AB})=S(\rho_{A})-\min_{\{P(\pm x)\}} \{p(x) S(\rho_{A|P^{B}(x)}) + p(-x) S(\rho_{A|P^{B}(-x)})\}. \nonumber
\end{eqnarray}

As weak measurements disturb the subsystem of a composite system weakly, the information is less lost and destroyed by weak measurements on the subsystem alone.
That is the physical interpretation that the super classical correlation is smaller than the classical correlation, $J_{B}^{w}(C_{1}^{w})\leq J_{B}(C_{1})$.
According this fact, we can infer that weak measurements can capture more quantum correlation than the projection measurements.
In fact, the super quantum correlation $D_w(\rho_{AB})=I(\rho_{AB})-J_B^{w}(\rho_{AB})$ is lager than the quantum correlation $D(\rho_{AB})=I(\rho_{AB})-J_B(\rho_{AB})$.
And there is a similarity to the Holevo quantity which measures the capacity of quantum states for classical communication.

\subsection{Dynamics of weak maximal Holevo quantity of Bell-diagonal states under local nondissipative channels.}
We will consider the system-environment interaction\cite{nielsen} through the evolution of a quantum state
$\rho$ under a trace-preserving quantum operation $\varepsilon(\rho)$,
\begin{equation}
\varepsilon(\rho) = \sum_{i,j} \left(E_i\otimes E_j\right) \rho \left(E_i \otimes E_j\right)^\dagger,\nonumber
\end{equation}
where $\{E_k\}$ is the set of Kraus operators associated to a decohering process of a single qubit,
with $\sum_k E_k^\dagger E_k = I$. We will use the Kraus operators in Table~\ref{t1} \cite{mont} to describe a variety of channels considered in this work.

\begin{table}[hbt]
\begin{center}
\begin{tabular}{|c|c|}
\hline
 & $\textrm{Kraus operators}$                                         \\ \hline \hline
 & \\
BF   & $E_0 = \sqrt{1-p/2}\, I , E_1 = \sqrt{p/2} \,\sigma_1$                        \\ \hline
 & \\
PF   & $E_0 = \sqrt{1-p/2}\, I , E_1 = \sqrt{p/2}\, \sigma_3$                        \\ \hline
 & \\
BPF & $E_0 = \sqrt{1-p/2}\, I , E_1 = \sqrt{p/2} \,\sigma_2$                        \\ \hline
 & \\
GAD   &
$E_0=\sqrt{p}\left(
\begin{array}{cc}
1 & 0 \\
0 & \sqrt{1-\gamma} \\
\end{array} \right) ,
E_2=\sqrt{1-p}\left(
\begin{array}{cc}
\sqrt{1-\gamma} & 0 \\
0 & 1 \\
\end{array} \right)$  \\
& \\
 & $E_1=\sqrt{p}\left(
\begin{array}{cc}
0 & \sqrt{\gamma} \\
0 & 0 \\
\end{array} \right) ,
E_3=\sqrt{1-p}\left(
\begin{array}{cc}
0 & 0 \\
\sqrt{\gamma} & 0 \\
\end{array} \right)$  \\ \hline
\end{tabular}
\caption[table1]{Kraus operators for the quantum channels: bit flip (BF), phase flip (PF),
bit-phase flip (BPF), and generalized amplitude damping (GAD), where
$p$ and $\gamma$ are decoherence probabilities, $0<p<1$, $0<\gamma<1$.}
\label{t1}
\end{center}
\end{table}

The decoherence processes BF, PF, and BPF in Table~\ref{t1} preserve the Bell-diagonal form of the
density operator $\rho_{AB}$.
For the case of GAD, the Bell-diagonal form is kept for arbitrary $\gamma$
and $p=1/2$. In this situation, we can write the quantum operation $\varepsilon(\rho)$ as
\begin{equation}
\varepsilon(\rho_{AB})=\frac{1}{4}(I\otimes I+\sum_{i=1}^3c^\prime_i\sigma_i\otimes\sigma_i)\label{newstate},
\end{equation}
where the values of the $c^\prime_1$, $c^\prime_2$, $c^\prime_3$
are given in Table~\ref{t2} \cite{mont}.
\begin{table}[hbt]
\begin{center}
\begin{tabular}{|c|c|c|c|}
\hline
$\textrm{Channel}$ & $c^\prime_1$      & $c^\prime_2$     & $c^\prime_3$      \\ \hline \hline
& & & \\
BF                 &  $c_1$            & $c_2 (1-p)^2$    & $c_3 (1-p)^2$     \\ \hline
& & & \\
PF                 &  $c_1 (1-p)^2$    & $c_2 (1-p)^2$    & $c_3$             \\ \hline
& & & \\
BPF                &  $c_1 (1-p)^2$    & $c_2$            & $c_3 (1-p)^2$     \\ \hline
& & & \\
GAD                &  $c_1 (1-\gamma)$ & $c_2 (1-\gamma)$ & $c_3 (1-\gamma)^2$ \\ \hline
\end{tabular}
\caption[table2]{Correlation functions for the quantum operations: bit flip (BF), phase flip (PF),
bit-phase flip (BPF), and generalized amplitude damping (GAD). For GAD, we
fixed $p=1/2$.}
\label{t2}
\end{center}
\end{table}

 When $|c_{1}|=\max\{|c_{1}|, |c_{2}|, |c_{3}|\}$, $|c_{3}|=\max\{|c_{1}|, |c_{2}|, |c_{3}|\}$, $|c_{2}|=\max\{|c_{1}|, |c_{2}|, |c_{3}|\}$,  respectively, we have that $|c_{1}|, |c_{3}|, |c_{2}|$ are the maximal values among $c^\prime_1$, $c^\prime_2$, $c^\prime_3$ in each line of Tabel~\ref{t2} .
 As $\varepsilon(\rho)$ is also Bell-diagonal states, from Eqs. (\ref{max}), (\ref{cc}), (\ref{jc}), (\ref{sc}), (\ref{sjc}) we find that all of the classical correlation, the maximal Holevo quantity, the super classical correlation and the weak maximal Holevo quantity for Bell-diagonal states through any channel of bit flip, phase flip, bit-phase flip remain unchanged.
 In particular, for Werner states, we find that all the classical correlation, the maximal Holevo quantity, the super classical correlation and the weak maximal Holevo quantity for Werner states keep unchanged under all the channel of bit flip, phase flip, bit-phase flip.

The maximal Holevo quantity of the Werner states under generalized amplitude damping is given by
\begin{eqnarray}
NC_{1}(\rho_{AB})&=&\frac{1-z(1-\gamma)}{2}\log[1-z(1-\gamma)]\nonumber\\
& &+\frac{1+z(1-\gamma)}{2}\log[1+z(1-\gamma)].
\end{eqnarray}

The weak maximal Holevo quantity of the Werner states under generalized amplitude damping is given by
\begin{eqnarray}
NC_{1}^{w}(\rho_{AB})&=&\frac{1-z(1-\gamma)\tanh x}{2}\log[1-z(1-\gamma)\tanh x]\nonumber\\
& &+\frac{1+z(1-\gamma)\tanh x}{2}\log[1+z(1-\gamma)\tanh x].
\end{eqnarray}

In Fig.2, as an example, the dynamic behaviors of the weak maximal Holevo quantity and the maximal Holevo quantity of the Werner  states under the generalized amplitude damping channel are depicted for $x=0.5$ and $x=1$.
 Against the decoherence, when $x$ increases, the weak maximal Holevo quantity become greater.
 The weak maximal Holevo quantity approaches to the maximal Holevo quantity for larger $x$ under the generalized amplitude damping channel.
 The weak maximal Holevo quantity and the maximal Holevo quantity increase as $z$ increases.
 Then as $\gamma$ increases, the weak maximal Holevo quantity and the maximal Holevo quantity decrease.

\subsection{Weak measurements can be realized by noise and project measurements}
Now we study the realization of weak measurements by means of depolarizing noise and project measurements.
The depolarizing noise is an important type of quantum noise that transforms a single qubit state into a completely mixed state $I/2$ with probability $p$ and leaves a qubit state untouched with probability $1-p$.
The operators for single qubit depolarizing noise are given by \cite{jia}
\begin{eqnarray*}
D_{1}&=& \sqrt{1-p}
\left(\begin{array}{rr}
1 & 0  \\  0 &  1
\end{array} \right)
, \ \
D_{2}= \sqrt{\frac{p}{3}}
\left(\begin{array}{rr}
0 & 1  \\  1 & 0
\end{array} \right), \nonumber \\
D_{3}&=&\sqrt{\frac{p}{3}}
\left(\begin{array}{rr}
0 & -i  \\  i &  0
\end{array} \right), \ \
D_{4}=\sqrt{\frac{p}{3}}
\left(\begin{array}{rr}
1 & 0  \\  0 &  -1
\end{array} \right),
\end{eqnarray*}
where ~$ p = 1-e^{-\tau t}$. Then the Bell-diagonal states under the depolarizing noise acting on the first qubit of quantum state $\rho_{AB}$ are given by\cite{jia}
\begin{eqnarray}
\varepsilon(\rho_{AB})=\frac{1}{4}\left[I\otimes I+(1-\frac{4p}{3})\sum_{i=1}^3c_i\sigma_i\otimes\sigma_i\right].
\end{eqnarray}

As $\varepsilon(\rho_{AB})$ is also a Bell-diagonal state, after projective measurements on $B$, see Eq. (\ref{project}) in section Method,
the state $\varepsilon(\rho_{AB})$ becomes the following ensemble with
 $p_{0}=p_{1}=\frac{1}{2}$ and
\begin{eqnarray}\label{project2}
\rho_{0}=\frac{1}{2}\left[I+(1-\frac{4p}{3})(c_{1}z_{1}\sigma_{1}+c_{2}z_{2}\sigma_{2}+c_{3}z_{3}\sigma_{3})\right],\nonumber\\
\rho_{1}=\frac{1}{2}\left[I-(1-\frac{4p}{3})(c_{1}z_{1}\sigma_{1}+c_{2}z_{2}\sigma_{2}+c_{3}z_{3}\sigma_{3})\right].
\end{eqnarray}

Comparing Eq. (\ref{project2}) with the ensemble after weak measurements Eq. (\ref{wmea}) in section Method,
when $1-\frac{4p}{3}=\tanh x$, we obtain that weak measurements can be realized by means of depolarizing noise and projective measurements.

\section*{Discussion}

We have evaluated analytically the maximal Holevo quantity for Bell-diagonal states and find that it equals to the classical correlation.
We have given the definition of ``super classical correlation'' by weak measurements as the generalization of classical correlation defined by
standard projective measurements. We have evaluated super classical correlation for Bell-diagonal states and find that it is smaller than the  classical correlation and approaches the
classical correlation by tuning the weak measurements continuously to the projective measurements. We have shown the physical implications
that weak measurements can capture more quantum correlation than projective measurements.

As the generalization of the maximal Holevo quantity defined by projective measurements, we have also proposed
``weak maximal Holevo quantity'' by weak measurements. We have evaluated the weak maximal Holevo quantity for Bell-diagonal states and find
that it is smaller than the maximal Holevo quantity in general. Moreover, it has been shown that the weak maximal Holevo quantity equals to super classical correlation.

As applications, the dynamic behavior of the weak maximal Holevo quantity under decoherence has been investigated.
For some special Bell-diagonal states, we found that the weak maximal Holevo quantity remain unchanged under all the channels of bit flip, phase flip and bit-phase flip.

The dynamical behaviors of the weak maximal Holevo quantity for the Werner states under the generalized amplitude damping channel have been investigated.
Under the generalized amplitude damping channel, the weak maximal Holevo quantity becomes greater when $x$ increases and approaches to the maximal Holevo quantity for larger $x$.
The weak maximal Holevo quantity increases as $z$ increases. The weak maximal Holevo quantity decreases as $\gamma$ increases.
Above all, it has been shown that weak measurements can be realized by means of depolarizing noise and projective measurements.

The Holevo bound is a keystone in quantum information theory and plays important roles in many quantum information processing.
While the maximal Holevo quantity provides us different perspectives about classical correlations. The behaviors of weak maximal Holevo quantity vary a lot with the strength of the weak measurements.
Those measures can be applied to various protocols in quantum information processing, and identify the importance of the classical correlations in those protocols.

\section*{Methods}
\subsection{Calculation of the maximal Holevo quantity of Bell-diagonal states.}
We compute the maximal Holevo quantity $C_{1}(\rho_{AB})$ of Bell-diagonal states.
Let $\{\Pi_{k}=|k\rangle\langle k|, k=0, 1\}$ be the local measurements on the system $B$ along the computational base ${|k\rangle}$. Any von Neumann measurement on the system $B$ can be written as
\begin{eqnarray}
\{B_{k}=V\Pi_{k}V^{\dag}: k=0, 1\}\nonumber
\end{eqnarray}
for some unitary $V\in U(2)$. Any unitary $V$ can be written as
\begin{eqnarray}\label{V}
V=tI+i\vec{y}\vec{\sigma}
\end{eqnarray}
with $t\in R$, $\vec{y}=(y_{1}, y_{2}, y_{3})\in R^{3}$, and $t^{2}+y_{1}^{2}+y_{2}^{2}+y_{3}^{2}=1. $
After the measurements ${B_{k}}$, the state $\rho_{AB}$ will be changed to the ensemble $\{{\rho_{A\mid k}, p_{k}}\}$ with
\begin{eqnarray}
\rho_{A\mid k}: =\frac{1}{p_{k}}(I\otimes B_{k})\rho(I\otimes B_{k}),\nonumber\\
p_{k}=\mbox{tr}_B(I\otimes B_{k})\rho(I\otimes B_{k}).\nonumber
\end{eqnarray}

After some algebraic calculations\cite{luo}, we obtain $p_{0}=p_{1}=\frac{1}{2}$ and
\begin{eqnarray}
\rho_{A\mid0}=\frac{1}{2}\left[I+(c_{1}z_{1}\sigma_{1}+c_{2}z_{2}\sigma_{2}+c_{3}z_{3}\sigma_{3})\right],\nonumber\\
\rho_{A\mid1}=\frac{1}{2}\left[I-(c_{1}z_{1}\sigma_{1}+c_{2}z_{2}\sigma_{2}+c_{3}z_{3}\sigma_{3})\right],\label{project}
\end{eqnarray}
where
\begin{eqnarray}
z_{1}=2(-ty_{2}+y_{1}y_{3}), \quad z_{2}=2(ty_{1}+y_{2}y_{3}), \quad z_{3}=t^{2}+y_{3}^{2}-y_{1}^{2}-y_{2}^{2}.\label{condition6}\nonumber
\end{eqnarray}
Therefore,
\begin{eqnarray}
S(\sum_{i}p_{i}\rho_{A|i})=S(\frac{I}{2})=1.\label{con1}
\end{eqnarray}

Denote $\theta=\sqrt{|c_{1}z_{1}|^{2}+|c_{2}z_{2}|^{2}+|c_{3}z_{3}|^{2}}$. Then
\begin{eqnarray}
S(\rho_{A|0})=S(\rho_{A|1})=-\frac{1-\theta}{2}\log\frac{1-\theta}{2}-\frac{1+\theta}{2}\log\frac{1+\theta}{2},
\end{eqnarray}
and
\begin{eqnarray}
\sum_{i}p_{i}S(\rho_{A|i}))&=&\frac{1}{2}S(\rho_{A|0})+\frac{1}{2}S(\rho_{A|1})\nonumber\\
 &=& -\frac{1-\theta}{2}\log\frac{1-\theta}{2}-\frac{1+\theta}{2}\log\frac{1+\theta}{2}.
\end{eqnarray}

It can be directly verified that $z_{1}^{2}+z_{2}^{2}+z_{3}^{2}=1$.
Let
\begin{eqnarray}
C=\max\{|c_{1}|, |c_{2}|, |c_{3}|\},\label{max}
\end{eqnarray}
then we have $\theta\leq\sqrt{|C|^{2}(|z_{1}|^{2}+|z_{2}|^{2}+|z_{3}|^{2})}=C.$
Hence we get $\sup\limits_{\{V\}}\theta=C$ and $\theta\in[0, C]$.  It can be verified that $\sum_i p_i S(\rho_{A|i})$
is a monotonically decreasing function of $\theta$ in the interval of $[0, C]$.  The minimal value of $\sum_i p_i S(\rho_{A|i})$ can be attained at point $C$,

\begin{eqnarray}
\min_{\{\Pi_i^B\}} \sum_i p_i S(\rho_{A|i})= -\frac{1-C}{2}\log\frac{1-C}{2}-\frac{1+C}{2}\log\frac{1+C}{2}.\label{con2}
\end{eqnarray}

By Eqs. (\ref{con1}) and (\ref{con2}), we obtain

\begin{eqnarray}\label{cc}
C_{1}(\rho_{AB})&=&\max_{\{\Pi^B_i\}}\chi\{\rho_{AB}|\Pi^B_i\}\}.\nonumber\\
&=&1-\min_{\{\Pi_i^B\}} \sum_i p_i S(\rho_{A|i})\nonumber\\
&=&\frac{1-C}{2}\log(1-C)+\frac{1+C}{2}\log(1+C).
\end{eqnarray}
As $\rho_{A}=\frac{I}{2}$, the classical correlation $J_B(\rho_{AB})$ is given by
\begin{eqnarray}\label{jc}
J_B(\rho_{AB})&=&\sup_{\{B_k\}}\{S(\rho_{A})-\sum_i p_i S(\rho_{A|i})\}\nonumber \\
&=&S(\rho_{A})-\min_{\{\Pi_i^B\}} \sum_i p_i S(\rho_{A|i})\nonumber\\
&=&\frac{1-C}{2}\log(1-C)+\frac{1+C}{2}\log(1+C).
\end{eqnarray}

\subsection{Calculation of the weak maximal Holevo quantity of Bell-diagonal states.}

Let $\{\Pi_{k}=|k\rangle\langle k|, k=0, 1\}$
be the local measurements for the part $B$ along the computational base ${|k\rangle}$. Then any weak measurement operators on the system $B$ can be written as
\begin{eqnarray}
I\otimes P(\pm x)=\sqrt{\frac{(1\mp\tanh x)}{2}}I\otimes V\Pi_0V^{\dag} + \sqrt{\frac{(1\pm\tanh x)}{2}}I\otimes V\Pi_1 V^{\dag},
\end{eqnarray}
for some unitary $V\in U(2)$ of the form Eq. (\ref{V}).

After weak measurements the resulting ensemble is given by $\{p(\pm x),\rho_{A|P^{B}(\pm x)}\}$.
We need to evaluate $\rho_{A|P^{B}(\pm x)}$ and $p(\pm x)$. By using the relations \cite{luo},
\begin{eqnarray}
V^{\dag}\sigma_{1}V=(t^{2}+y_{1}^{2}-y_{2}^{2}-y_{3}^{2})\sigma_{1}+2(ty_{3}+y_{1}y_{2})
\sigma_{2}+2(-ty_{2}+y_{1}y_{3})\sigma_{3},\nonumber\\[2mm]
V^{\dag}\sigma_{2}V=2(-ty_{3}+y_{1}y_{2})\sigma_{1}+(t^{2}+y_{2}^{2}-y_{1}^{2}-y_{3}^{2})
\sigma_{2}+2(ty_{1}+y_{2}y_{3})\sigma_{3},\nonumber\\[2mm]
V^{\dag}\sigma_{3}V=2(ty_{2}+y_{1}y_{3})\sigma_{1}+2(-ty_{1}+y_{2}y_{3})
\sigma_{2}+(t^{2}+y_{3}^{2}-y_{1}^{2}-y_{2}^{2})\sigma_{3},\nonumber
\end{eqnarray}
and
$\Pi_{0}\sigma_{3}\Pi_{0}=\Pi_{0}$, $\Pi_{1}\sigma_{3}\Pi_{1}=-\Pi_{1}$, $\Pi_{j}\sigma_{k}\Pi_{j}=0$
for $j=0,1$, $k=1, 2$, from Eqs. (\ref{probab}) and (\ref{state1}), we obtain $p(\pm x) =\frac{1}{2}$ and
\begin{eqnarray}
\rho_{A|P^{B}(+x)}=\frac{1}{2}\left[I-\tanh x(c_{1}z_{1}\sigma_{1}+c_{2}z_{2}\sigma_{2}+c_{3}z_{3}\sigma_{3})\right],\nonumber\\
\rho_{A|P^{B}(-x)}=\frac{1}{2}\left[I+\tanh x(c_{1}z_{1}\sigma_{1}+c_{2}z_{2}\sigma_{2}+c_{3}z_{3}\sigma_{3})\right],\label{wmea}
\end{eqnarray}
where $z_{1}=2(-ty_{2}+y_{1}y_{3})$, $z_{2}=2(ty_{1}+y_{2}y_{3})$ and $z_{3}=t^{2}+y_{3}^{2}-y_{1}^{2}-y_{2}^{2}$.

Therefore, we see that
\begin{eqnarray}
 S(\sum_{\pm x}p(\pm x)\rho_{A|P^{B}(\pm x)})=S(\frac{I}{2})=1.\label{con10}
\end{eqnarray}

Denote $\theta=\sqrt{|c_{1}z_{1}|^{2}+|c_{2}z_{2}|^{2}+|c_{3}z_{3}|^{2}}$. Then
\begin{eqnarray}
S(\rho_{A|P^{B}(+x)})=S(\rho_{A|P^{B}(-x)})=-\frac{1-\theta\tanh x}{2}\log\frac{1-\theta\tanh x}{2}-\frac{1+\theta\tanh x}{2}\log\frac{1+\theta\tanh x}{2},
\end{eqnarray}
and
\begin{eqnarray}
 S_w(A|\{P^{B}(x)\})&=&\frac{1}{2}S(\rho_{A|P^{B}(x)})+\frac{1}{2}S(\rho_{A|P^{B}(-x)})\nonumber\\
 &=& -\frac{1-\theta\tanh x}{2}\log\frac{1-\theta\tanh x}{2}-\frac{1+\theta\tanh x}{2}\log\frac{1+\theta\tanh x}{2}.
\end{eqnarray}

Let $C=\max\{|c_{1}|, |c_{2}|, |c_{3}|\},$ then $\theta\leq\sqrt{|c|^{2}(|z_{1}|^{2}+|z_{2}|^{2}+|z_{3}|^{2})}=C$. Hence we get $\sup\limits_{\{V\}}\theta=C$
and $\theta$ is $\theta\in[0, C]$.  It can be verified that $S_w(A|\{P^{B}(x)\})$
is a monotonically decreasing function of $\theta$ in the interval of $[0,C]$.  The minimal value of $S_w(A|\{P^{B}(x)\}$ can be attained at point $C$,
\begin{equation}
\min_{\{\Pi_i^B\}}  S_w(A|\{P^{B}(x)\})=-\frac{1-C\tanh x}{2}\log\frac{1-C\tanh x}{2}-\frac{1+C\tanh x}{2}\log\frac{1+C\tanh x}{2}.\label{con20}
\end{equation}

By Eqs. (\ref{con10}) and (\ref{con20}), we obtain
\begin{eqnarray}\label{sc}
C_{1}^{w}(\rho_{AB})&=& \max_{\{P(\pm x)\}}\chi^{w}\{\rho_{AB}|\{P(\pm x)\}\}\nonumber\\
&=&1-\min_{\{P(\pm x)\}}  S_w(A|\{P^{B}(x)\})\nonumber\\
&=&\frac{1-C\tanh x}{2}\log(1-C\tanh x)+\frac{1+C\tanh x}{2}\log(1+C\tanh x).
\end{eqnarray}

As $\rho_{A}=\frac{I}{2}$, the super classical correlation $J_B^{w}(\rho_{AB})$ is given by
\begin{eqnarray}\label{sjc}
J_B^{w}(\rho_{AB})&=&\sup_{\{B_k\}}\{S(\rho_{A})-S_w(A|\{P^{B}(x)\})\} \nonumber\\
&=&S(\rho_{A})-\min_{\{P(\pm x)\}} \{p(x) S(\rho_{A|P^{B}(x)}) + p(-x) S(\rho_{A|P^{B}(-x)})\}\nonumber\\
&=&\frac{1-C\tanh x}{2}\log(1-C\tanh x)+\frac{1+C\tanh x}{2}\log(1+C\tanh x).
\end{eqnarray}

\begin{addendum}

\item [Acknowledgement]

This work was supported by the National Natural Science Foundation of China under
grant No.

\item [Author Contributions]
Y.-K. W., S. -M. F., Z.-X. W, J.-P. C. and H. F. calculated and analyzed
the results. Y.-K. W. and H. F. co-wrote the paper. All authors
reviewed the manuscript and agreed with the submission.
\item [Competing Interests]
The authors declare that they have no competing financial interests.

\item [Correspondence]
Correspondence and requests for materials should be addressed to
H.F. or Y.-K. W.
\end{addendum}
\clearpage

\newpage

\bigskip

\textbf{Figure 1 Weak maximal Holevo quantity(super classical correlation) (dashed green line), maximal Holevo quantity(classical correlation) (solid blue line), quantum discord(solid cyan line), super quantum discord (dashed black line), and entanglement of formation(solid red line) for the Werner states as a function of $z$:  $x=0.25$ and $x=2.5$.}

\bigskip

\textbf{Figure 2 The weak maximal Holevo quantity(super classical correlation)$\{$$x=0.5$(blue surface), $x=1$(gray surface)$\}$ and the maximal Holevo quantity(classical correlation)(orange surface) for the Werner states under generalized amplitude damping channel as a function of $z$ and $\gamma$.}

\newpage

\begin{figure}[tbp]
\begin{center}
\epsfig{file=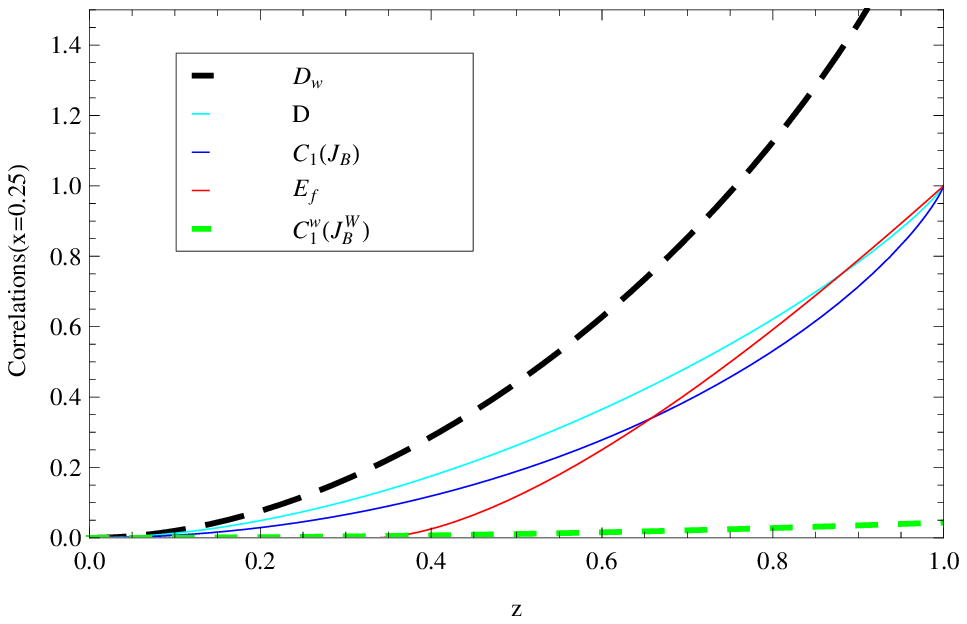, width=15cm}
\epsfig{file=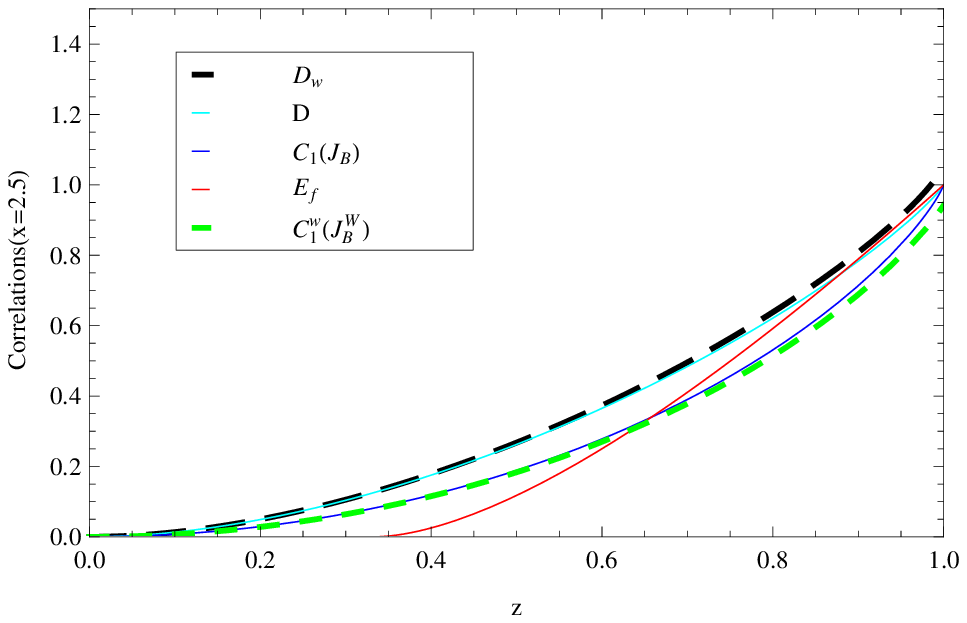, width=15cm}
\end{center}
\par
\label{fig1}
\end{figure}

\newpage

\begin{figure}[tbp]
\begin{center}
\epsfig{file=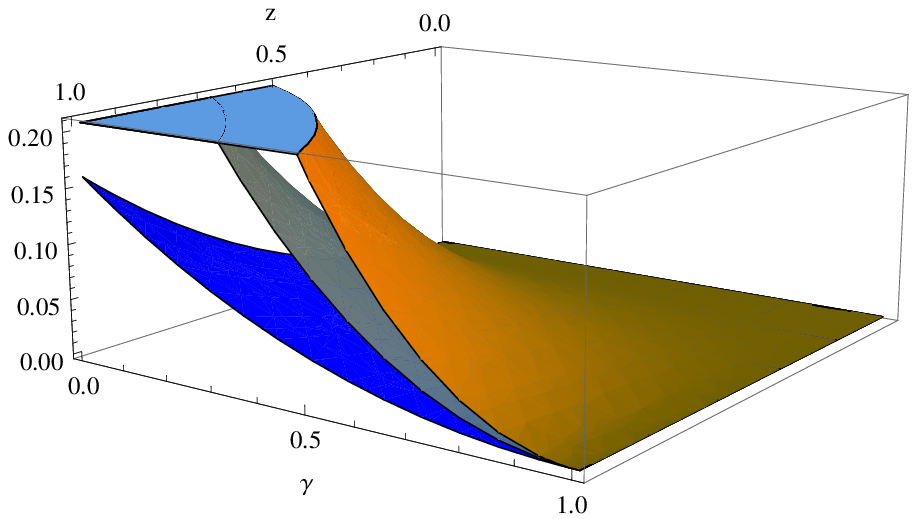, width=15cm}
\end{center}
\par
\label{fig2}
\end{figure}

\end{document}